\documentclass[]{JHEP3}
\usepackage{epsf,epsfig,subfigure,axodraw,graphicx,amsmath,amssymb}

\title{\Large\bf PAMELA/ATIC Anomaly from \\
Exotic Mediated Dark Matter Decay }

\author{\large
Kyu Jung Bae and Bumseok Kyae\\
Department of Physics and Astronomy and Center for
Theoretical Physics, Seoul National University, Seoul 151-747\\
E-mail: \email{baekj81@phya.snu.ac.kr, bkyae@phya.snu.ac.kr} }

\abstract{

We discuss dark matter decay mediated by exotically charged
particles (``exotics'') in a supersymmetric model with two dark
matter (DM) components: One is the (bino-like) lightest
supersymmetric particle (LSP) $\chi$, and the other is a newly
introduced meta-stable neutral singlet $N$.  $N$ decays to $\chi
e^+e^-$ via a dimension 6 operator induced by a penguin-type one
loop diagram with the life time of $10^{26}$ sec., explaining
energetic cosmic $e^\pm$ excess observed recently by PAMELA and
ATIC/PPB-BETS.  The superheavy masses of exotics ($\sim
10^{15-16}$ GeV) are responsible for the longevity of $N$. The
superpartner of $N$ develops the vacuum expectation value (VEV) of
order TeV so that the DM $N$ achieves the desired mass of 2 TeV.
By the VEV, the U(1)$_{\rm R}$ symmetry is broken to the discrete
$Z_2$ symmetry, which is identified with the matter parity in the
minimal supersymmetric standard model (MSSM). Since we have the
two DM components, even extremely small amount of $N$ [${\cal
O}(10^{-10})\lesssim (n_N/n_\chi)$] could account for the observed
positron flux with relatively light exotics' masses [$10^{12}~{\rm
GeV}\lesssim M_{\rm exo.}\lesssim 10^{16}~{\rm GeV}$]. }

 \preprint{SNUTP 09-004}

\begin{document}


\section{Introduction}

The recently reported observations by PAMELA
\cite{PAMELAe,PAMELAp} and ATIC/PPB-BETS \cite{ATIC,PPB-BETS}
collaborations on excess of high energy positrons from cosmic ray
have attracted more and more attentions. As many literatures
pointed out, dark matter (DM) decay \cite{strumia,decay,Ndecay} or
annihilation \cite{annihil} would be deeply involved in the
observed positron excess.\footnote{Alternatively, astrophysical
sources such as pulsars could explain the positron excess
\cite{pulsar}.} If it is indeed caused by DM, however, the
observation should be accepted as a puzzle, because it is hard to
be understood within the framework of the conventional DM
scenario, particularly, by the minimal supersymmetric standard
model (MSSM). Thus, the observed positron excess might be a hint
toward a new physics beyond the standard model (SM).

As noticed in Refs.~\cite{strumia}, the positron flux needed to
explain the observation of ATIC/PPB-BETS (and also PAMELA) can be
produced by leptonic decay of DM \cite{decay,Ndecay} with 2 TeV
mass ($=m_{\rm DM}$) via a dimension 6 operator (four fermion
interaction) suppressed by $M_{\rm GUT}^2\sim (10^{16}~{\rm
GeV})^2$, by which the decay rate is estimated as
\begin{eqnarray} \label{decayrate}
\Gamma_{\rm DM} \sim \frac{m_{\rm DM}^5}{192\pi^3M_{\rm
GUT}^4}\sim 10^{-26}~{\rm sec.}^{-1}  .
\end{eqnarray}
Hadronic decay channels should not exceed 10 $\%$ to be consistent
with the PAMELA's data \cite{PAMELAp}. The DM decay scenario
avoids the constraint from the $\gamma$ ray flux
\cite{aticGammaray} by the HESS observations of galactic ridge
\cite{HESS}. However, it is not trivial to see which physics at
the $M_{\rm GUT}$ scale can provide such a low energy effective
four fermion interaction, allowing DM to decay dominantly into the
SM leptons: In most of grand unified theories (GUTs) embedding the
SM, the gauge interactions by superheavy gauge boson exchanges can
easily provide four fermion interactions suppressed by $M_{\rm
GUT}^2$.  But they do not prefer only such a leptophilic decay
mode of an electrically neutral particle or DM. Hence, one should
explore the possibility of a leptophilic ${\it Yukawa}$
interaction for DM decay.

Recently, supersymmetric (SUSY) models possessing one more dark
matter component $N$ apart from the (bino-like) lightest
supersymmetric particle (LSP) $\chi$ have been suggested as the
resolutions of the PAMELA/ATIC anomaly \cite{HKK,BHKKV,Ndecay}.
Particularly in the model of Ref.~\cite{Ndecay}, the anomaly is
explained by decay of the extra DM component $N$ into $\chi
e^+e^-$ through a dimension 6 operator. The effective dimension 6
operator for DM decay is obtained from some renormalizable
leptophilic Yukawa interactions with the dimensionless coupling of
order unity, after a pair of vector-like SU(2) lepton doublets
$(L,L^c)$ and lepton singlets $(E,E^c)$ decoupled. The superheavy
masses of $L^{(c)}$, $E^{(c)}$ ($\sim 10^{16}$ GeV) are
responsible for the longevity of $N$. Since the gauge group is
just that of the SM and the low energy field spectrum is the same
as that of the MSSM except the neutral singlet $N$, the gauge
coupling unification in the MSSM is protected in the model. This
model is easily embedded in flipped SU(5), which is a leptophilic
unified theory \cite{BHKKV}.

Most of phenomenologically promising string models predict a lot
of vector-like {\it superheavy} exotic states (``exotics'')
carrying fractional electric charges \cite{stringMSSM}. One might
expect that such superheavy exotics also can play the role of
$(L,L^{c})$ and $(E,E^{c})$ in the model of Ref.~\cite{Ndecay},
mediating DM decay via the dimension 6 process. Their superheavy
masses ($\sim 10^{16}$ GeV) could lead successfully to $10^{26}$
sec. life time of the DM as desired. Considering the case that
superheavy exotics mediate DM decay, however, one should notice a
remarkable point: Most of all, fractionally charged heavy
particles can not decay to the light SM leptons, because of the
charge conservation. Thus, if exotics are involved in the process,
$N\rightarrow e^++e^-+{\it neutral~particles}$, where the initial
and final states are the states only with the integral electric
charges, they should be co-created and co-annihilated between the
initial and final states. It means that DM decay is possible only
at loop levels, if exotics dominantly mediate DM decay.

In this paper, we explore the possibility that DM decay is
mediated by a one loop diagram.  If the mediators are indeed
fractionally charged superheavy particles, we should necessarily
consider the loop induced process.  However, our study is not
confined only to the case of fractionally charged heavy field
mediation, but covers more general cases of loop induced DM
decays.

\section{The model}

Let us consider the vector-like superheavy superfields
$(E,E^{c})$, $(X,X^{c})$, and $(O,O^{c})$.  Their quantum numbers
are shown in Table 1.
If $q$ is a fractional number, $E^{(c)}$, $X^{(c)}$, and $O^{(c)}$
become regarded as exotics. In Table 1, we present only the first
generation of the charged lepton singlets, $e^c$. Concerning the R
charges of the other MSSM superfields, we assign $1$ to the MSSM
matter superfields like $e^c$, and $0$ to the two MSSM Higgs
doublets. We leave open the possibility that $E^{(c)}$, $X^{(c)}$,
and $O^{(c)}$ are charged also under other (visible or hidden
gauge) symmetry ${\cal G}$.  For the case that this model is
embedded in flipped SU(5) [$=$SU(5)$\times$U(1)$_X$], ${\cal G}$
can correspond to SU(5).

\begin{table}[!h]
\begin{center}
\begin{tabular}
{c|cccccccc} {\rm Superfields} ~\quad & ~\quad  $e^c$  ~\quad &
~\quad $N$ ~\quad & ~\quad $E$ ~\quad & ~\quad $E^c$ ~\quad &
~\quad $X$ ~\quad & ~\quad $X^c$ ~\quad &
$O$ & $O^c$ \\
\hline U(1)$_{\rm Y}$ & $1$ & $0$ & $q$ & $-q$ & $-q$ & $q$ &
~$q-1$ ~& $-q+1$
\\
U(1)$_{\rm R}$ & 1 & 2/3 & 1/3 & 5/3
 & 1 & 1 & 0 & 2 \\
(~${\cal G}$~)~ & ${\bf 1}$ & ${\bf 1}$ & (~${\cal R}$~) &
(~${\cal R}^*$~) & (~${\cal R}^*$~) & (~${\cal R}$~) & (~${\cal
R}$~) & (~${\cal R}^*$~)
\end{tabular}
\end{center}\caption{The hypercharges and R charges of the superfields.
The hypercharge $q$ can be a fractional number.  The vector-like
exotic superfields, $E^{(c)}$, $X^{(c)}$, and $O^{(c)}$ are all
decoupled from low energy physics due to their heavy masses. The
(visible or hidden) symmetry ${\cal G}$ is optional.
}\label{tab:Qnumb}
\end{table}
%
%

If the deviation of $(e^++e^-)$ observed by ATIC/PPB-BETS from
cosmic ray is indeed caused by DM decay, the mass of DM should be
around 2 TeV \cite{strumia}. In order to protect the status of
SUSY as the solution of the gauge hierarchy problem, we should
assume that the mass of the LSP is of ${\cal O}(100)$ GeV or
lighter. Apart from the (bino-like) LSP $\chi$, thus, we introduce
one more dark matter component with 2 TeV mass, which is the
fermionic component of $N$ in Table 1, to account for the
ATIC/PPB-BETS' data.

The relevant superpotential in our model is composed of the
trilinear and bilinear terms: $W=W_{{\rm tri}}+W_{\rm bi}$, where
$W_{{\rm tri}}$ and $W_{\rm bi}$ are, respectively, given by
\begin{eqnarray} \label{WNdecay}
W_{{\rm tri}} &=& NEX + XOe^c + N^3  ,
\\ \label{Wmass}
W_{\rm bi} &=& M_EEE^c + M_X XX^c + M_O OO^c .
\end{eqnarray}
We dropped the dimensionless Yukawa coupling constants in
Eq.~(\ref{WNdecay}) for simplicity.  They are tacitly assumed to
be {\it of order unity}. The dimensionful parameters, $M_E$,
$M_X$, and $M_O$ in Eq.~(\ref{Wmass}) are $10^{15}$--$10^{16}$
GeV. Thus, the vector-like fields $(E, E^c)$, $(X, X^c)$, and $(O,
O^c)$ are superheavy. To avoid couplings with the other charged
lepton singlets, $\mu^c$ and $\tau^c$, one can introduce a family
dependent U(1)$_{\rm PQ}$ symmetry.  It can explain the smallness
of the electron mass \cite{BHKKV,Ndecay}.  The $N^3$ term in
Eq.~(\ref{WNdecay}) is introduced such that the scalar component
of $N$, i.e. $\tilde{N}$ promptly decays into the two fermionic
components  $2N$:
The mass of the fermionic component of $N$ ($\approx 2$ TeV) is
induced by the vacuum expectation value (VEV) $\langle
\tilde{N}\rangle$. On the other hand, the mass squared of the
scalar component of $N$ is given by $(|\langle
\tilde{N}\rangle|^2+m_{3/2}^2)$, where $m_{3/2}^2$ comes from the
soft scalar mass term of $\tilde{N}$. We will discuss later how
the VEV of $\tilde{N}$ could be developed.
We just assume that the soft mass of $\tilde{N}$ is heavy enough
($\gtrsim 4$ TeV) for the decay $\tilde{N}\rightarrow 2N$ to be
possible. Since we don't want the $N^2$ term with a too large mass
parameter in the superpotential, we employ the U(1)$_{\rm R}$
symmetry to forbid it from the bare superpotential.

This model is easily embedded in flipped SU(5) \cite{Barr82}. To
account for the PAMELA's important observation, i.e. no excess of
anti-proton \cite{PAMELAp}, the lepton singlet $e^c$ should not be
accompanied with quarks in Eq.~(\ref{WNdecay}), when the model
embedded in a GUT. Since in flipped SU(5) $e^c$ and $N$ remain
SU(5) singlets, ${\bf 1_5}$ and ${\bf 1_0}$, respectively, flipped
SU(5) models can be perfectly consistent with the PAMELA's data
\cite{BHKKV}. Moreover, flipped SU(5) is phenomenologically
attractive: The notorious doublet/triplet splitting problem in
GUTs is very easily resolved via the missing partner mechanism
\cite{Barr82}. The predicted fermion mass relation in flipped
SU(5) is just that between up-type quarks and Dirac neutrinos
masses. Since the Majorana neutrino masses are still not
constrained, however, the mass relation in flipped SU(5) does not
encounter any difficulty in matching the real data on fermion
masses.

The presence of the A-term corresponding to $N^3$, i.e.
$(m_{3/2}'\tilde{N}^3+{\rm h.c.})$, and $|\tilde{N}|^4$, (and also
the soft mass term $m_{3/2}^2|\tilde{N}|^2$) in the scalar
potential permits two vacua, on which $\langle\tilde{N}\rangle=0$
and $\langle\tilde{N}\rangle\sim {\cal O}(m_{3/2})$, respectively.
We assume that our universe is at the latter, which can be the
absolute minimum of the scalar potential for a proper set of the
parameters. Then, the Majorana mass term of the fermionic
component of $N$, i.e. $m_{N}N^2$ is generated in the
superpotential:
\begin{eqnarray}
\langle \tilde{N}\rangle ~\sim ~ m_{N} ~\sim ~ {\cal O}(m_{3/2}) .
\end{eqnarray}
Since we regard the fermionic component of $N$ as the extra DM
component explaining the ATIC/PPB-BETS' observation, we take
$m_N=m_{\rm DM}\approx 2$ TeV.

The non-vanishing VEV $\langle\tilde{N}\rangle$ breaks U(1)$_{\rm
R}$ to the discrete $Z_2$ symmetry, because the unit R charge is
$1/3$ in this model. Since the superfields carrying ${\rm
R}=1/3,1,5/3$ ($0,2/3,2$) become odd (even) under $Z_2$, the
remaining $Z_2$ symmetry is exactly identified with the R (or
matter) parity. In fact, the U(1)$_{\rm R}$ breaking source is the
SUSY breaking source $\langle F\rangle\sim m_{3/2}M_P\sim
(10^{10}~{\rm GeV})^2$, which is the VEV of the F-component of a
hidden sector superfield, and generates the SUSY breaking soft
terms in the visible sector. Since the R parity of $N$ is even,
$N$ can not be the Majorana neutrino participating in the seesaw
mechanism. Since the coupling of $N$ to the MSSM Higgs doublets is
possible, at best, only at the high order superpotential,
$(\langle\tilde{N}\rangle^2/M_P^2)Nh_uh_d$, it can not also be the
extra singlet appearing in the ``next-to-minimal supersymmetric
standard model (NMSSM)'' \cite{NMSSM}.

With the terms in the superpotential Eq.~(\ref{WNdecay}), the DM
$N$ can decays to $\chi e^+e^-$ via a dimension 6 operator induced
by a one loop diagram, if $m_{\rm DM}\lesssim m_{\tilde{e}^c}$:
\begin{eqnarray}
N \longrightarrow \chi+e^-+e^+  .
\end{eqnarray}
See the dominant Feynman diagram in Figure 1, which looks similar
to ``Penguin diagram'' appearing in the K and B meson decays.
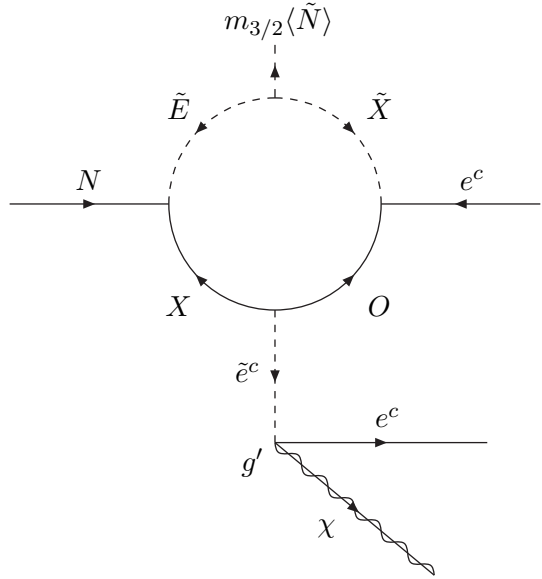
\begin{figure}[t]
\begin{center}
\begin{picture}(240,280)(0,60)

\DashArrowArc(120,200)(40,90,180){3}
\DashArrowArcn(120,200)(40,90,0){3}
\ArrowArcn(120,200)(40,270,180)\ArrowArc(120,200)(40,270,360)

\ArrowLine(20,200)(80,200) \ArrowLine(220,200)(160,200)
\DashArrowLine(120,240)(120,260){3}
\DashArrowLine(120,160)(120,110){3}

\ArrowLine(120,110)(200,110) \ArrowLine(120,110)(180,60)
\Photon(120,110)(180,60){-2}{6}

\Text(120,270)[]{$~m_{3/2}\langle\tilde{N}\rangle$}
\Text(50,210)[]{$N$} \Text(195,210)[]{$e^c$}
\Text(110,137)[]{$\tilde{e}^c$} \Text(163,120)[]{$e^c$}
\Text(140,78)[]{$\chi$}

\Text(84,162)[]{$X$} \Text(84,238)[]{$\tilde{E}$}
\Text(160,238)[]{$\tilde{X}$} \Text(160,162)[]{$O$}
\Text(112,103)[]{$g'$}

\end{picture}
\caption{Penguin-type one loop decay diagram of $N$: It is the
dominant diagram of $N \to \chi+e^-+e^+$. The dimensionless Yukawa
couplings are of order unity.}\label{fig:penguin}
\end{center}
\end{figure}
As seen in Figure 1, the effective dimensionless coupling of
$\tilde{e}^{c*}Ne^c$ in the Lagrangian is induced by the loop. It
is estimated as
\begin{eqnarray}  \label{amp}
\frac{m_{3/2}\langle
\tilde{N}\rangle}{48\pi^2M_*^2} \times {\cal O}(y^4) \times {\cal
N} ,
\end{eqnarray}
where we set $M_{\tilde{E}}^2=M_X^2=M_{\tilde{X}}^2=M_O^2\equiv
M_*^2$. ${\cal O}(y^4)$ denotes the contributions of the
dimensionless Yukawa coupling constants, which are assumed to be
of order unity at the GUT scale. Since the superheavy fields are
involved in the relevant Yukawa terms in Eq.~(\ref{WNdecay}), the
couplings of the terms ``$NEX$,'' and ``$XOe^c$'' do not much
evolve with energy after the superheavy fields decoupled.
[The order of magnitudes of the $N^3$ coupling at the GUT and
lower energies are the same because of the small beta function
coefficient.]
Thus, the low energy effective coupling, i.e. Eq.~(\ref{amp}),
which is obtained by integrating out the superheavy particles, is
extremely small [$<{\cal O}(m_{3/2}^2/M_*^2)$].  If $E$, $X$, and
$O$ are in large dimensional representations under the other
(visible or hidden) non-abelian (gauge) groups ${\cal G}$, the
dimension ``${\cal N}$'' can be crucial in Eq.~(\ref{amp}). The
decay rate of $N \rightarrow \chi+e^-+e^+$ is estimated as
\begin{eqnarray}
\Gamma_N \approx \frac{m_{\rm DM}^5}{192\pi^3}\times
\left[\frac{g^{'}m_{3/2}\langle\tilde{N}\rangle}{~96\pi^2M_*^2m_{\tilde{e}^c}^2~}
\right]^2 \times {\cal O}(y^8) \times {\cal N}^2 ,
\end{eqnarray}
where $\Gamma_N\sim 10^{-26}~{\rm sec.}^{-1}$ for $m_{\rm DM}\sim
2$ TeV [$\gtrsim 10\times {\cal O}(m_\chi)$], $M_*\sim 10^{15}$
GeV, ${\cal O}(y^8)\sim 1$, and ${\cal N}=1$.  Note that if the
dimensionless Yukawa couplings in Eq.~(\ref{WNdecay}) are about 3,
$M_*$ can be slightly heavier upto $10^{16}$ GeV, yielding the
same decay rate.
The other non-abelian (global or gauge) symmetry ${\cal G}$, under
which $E$, $X$, and $O$ are charged, would be useful in raising
$M_*$ higher. For instance, if ${\cal G}=$ flipped SU(5) in the
visible sector and the superheavy fields are of the SU(5) tensor
representation, ${\cal R}={\bf 10}$ [or ${\cal G}=$ SO(10) in the
hidden sector and the superheavy fields are of the SO(10) vector
representations, ${\cal R}={\cal R }^*={\bf 10}$], then the
circulating fields on the loop are 10 times more (i.e. ${\cal
N}=10$) and so the decay rate is 100 times enhanced, compared to
the case of the singlets.

If the selectron $\tilde{e}^c$ is relatively light, $m_{\rm
DM}\gtrsim m_{\tilde{e}^c}$, then $\tilde{e}^c$ can be an on-shell
particle in Figure 1, and so the two body decay channel,
$N\rightarrow e^-+\tilde{e}^c$ opens. Thus, the decay rate becomes
enhanced by ${\cal O}(100)$:
\begin{eqnarray}
\Gamma_N\approx\frac{(m_{\rm DM}^2-m_{\tilde{e}^c}^2)^2}{16\pi
~m_{\rm DM}^3}\left[\frac{m_{3/2}\langle
\tilde{N}\rangle}{48\pi^2M_*^2}\right]^2\times {\cal O}(y^8)\times
{\cal N}^2.
\end{eqnarray}
For $\Gamma_N$ giving $10^{-26}$ sec$^{-1}$, thus, $M_*\sim
10^{15-16}$ GeV is not much affected.

Note that in this model, (anti-) neutrinos and charged leptons
heavier than the electron are not produced at all from the DM
decay. [The muons eventually decay to the electrons and (anti-)
neutrinos by the weak interaction.] Hence, this model is
completely free from the constraints on neutrino flux
\cite{superKK}.

In this model, we have the two DM components, $N$ and the
(bino-like) LSP $\chi$.  As noted in Ref.~\cite{Ndecay}, even
extremely small amount of $N$ [${\cal O}(10^{-10})\lesssim
(n_{N}/n_{\chi})$] can produce the positron flux needed to account
for PAMEL/ATIC data, only if the decay rate is enhanced by taking
relatively light masses of the exotic mediators [$10^{12}~{\rm
GeV}\lesssim M_*\lesssim 10^{16}~{\rm GeV}$]. Since the other DM
component, $\chi$ can still support the needed DM density
$\rho_{\rm DM}\approx 10^{-6}$ GeVcm$^{-3}$, thus, we have
extremely large flexibility for the portion of $n_N/n_\chi$.

We have already a TeV scale mass of $N$.  Thus, $N$ can play the
role of the well-known weakly interacting massive particle (WIMP)
such as the neutralino in the MSSM, e.g. if an interaction with
some other hidden sector fields $H$ and $H^c$, $W\supset y_hNHH^c$
is introduced. Here $y_h$ is a Yukawa coupling constant of order
unity and the masses of the scalar partners of $H$ and $H^c$ are
assumed to be of order the electroweak scale. [Then the
annihilation cross section of $N$ would be in the needed range for
explanation of dark matter ($\langle \sigma|v|\rangle\sim
10^{-27}$ cm$^3$s$^{-1}$).] $N$ could be in a thermal equilibrium
state with $H$, $H^c$ by exchanging their scalar partners down to
a proper decoupling temperature defined with hidden sector fields.
Departure of $N$ from the interactions could leave the relic
energy density of order $10^{-6}$ GeVcm$^{-3}$. Alternatively, $N$
could be non-thermally produced by decay of hidden sector fields.
However, we do not specify a possibility, because we have
extremely large flexibility of $n_N/n_\chi$.

\section{Conclusions}

Along the line of Ref.~\cite{Ndecay}, we proposed another SUSY
model with two DM components ($N,\chi$). A DM could decay to the
SM particles only at loop levels, when the exotics are the
mediator of the decay process. In this model, the extra DM
component $N$ decays to $\chi e^+e^-$ through a dimension 6
operator induced by a penguin-type one loop diagram. Its extremely
long life time $10^{26}$ sec. required for explaining the observed
positron excess is caused by the superheavy masses of exotic
states mediating the DM decay. Even with extremely small amount of
$N$, the positron excess could be explained.
This model is easily embedded in flipped SU(5), in which $e^c$ and
$N$ remain SU(5) singlets.

\acknowledgments{  We thank Jihn E. Kim for valuable discussions.
K.J.B. is supported in part by the FPRD of the BK21 program and
the Korea Science and Engineering Foundation grant funded by the
MEST through Center for Quantum Spacetime of Sogang University
with Grant Number R11-2005-021. B.K. is supported by the FPRD of
the BK21 program, in part by the Korea Research Foundation, Grant
No. KRF-2005-084-C00001 and the KICOS Grant No.
K20732000011-07A0700-01110 of the Ministry of Education and
Science of Republic of Korea.}

\newpage


\end{document}